\documentstyle{mn}

\edef\resetatcatcode{\catcode`\noexpand\@\the\catcode`\@\relax}
\ifx\miniltx\undefined\else \fi
\let\miniltx\box

\def\makeatletter{\catcode`\@11\relax}
\def\makeatother{\catcode`\@12\relax}
\makeatletter

\def\@makeother#1{\catcode`#1=12\relax}

\def\@ifnextchar#1#2#3{%
  \let\reserved@d=#1%
  \def\reserved@a{#2}\def\reserved@b{#3}%
  \futurelet\@let@token\@ifnch}
\def\@ifnch{%
  \ifx\@let@token\@sptoken
    \let\reserved@c\@xifnch
  \else
    \ifx\@let@token\reserved@d
      \let\reserved@c\reserved@a
    \else
      \let\reserved@c\reserved@b
    \fi
  \fi
  \reserved@c}
\begingroup
\def\:{\global\let\@sptoken= } \:  
\def\:{\@xifnch} \expandafter\gdef\: {\futurelet\@let@token\@ifnch}
\endgroup

\def\@ifstar#1{\@ifnextchar *{\@firstoftwo{#1}}}
\long\def\@dblarg#1{\@ifnextchar[{#1}{\@xdblarg{#1}}}
\long\def\@xdblarg#1#2{#1[{#2}]{#2}}

\long\def \@gobble #1{}
\long\def \@gobbletwo #1#2{}
\long\def \@gobblefour #1#2#3#4{}
\long\def\@firstofone#1{#1}
\long\def\@firstoftwo#1#2{#1}
\long\def\@secondoftwo#1#2{#2}

\def\NeedsTeXFormat#1{\@ifnextchar[\@needsf@rmat\relax}
\def\@needsf@rmat[#1]{}
\def\ProvidesPackage#1{\@ifnextchar[%
    {\@pr@videpackage{#1}}{\@pr@videpackage#1[]}}
\def\@pr@videpackage#1[#2]{\wlog{#1: #2}}

\let\DeclareOption\@gobbletwo

\def\RequirePackage{%
  \@fileswithoptions\@pkgextension}
\def\@fileswithoptions#1{%
  \@ifnextchar[
    {\@fileswith@ptions#1}%
    {\@fileswith@ptions#1[]}}
\def\@fileswith@ptions#1[#2]#3{%
  \@ifnextchar[
  {\@fileswith@pti@ns#1[#2]#3}%
  {\@fileswith@pti@ns#1[#2]#3[]}}

\def\@fileswith@pti@ns#1[#2]#3[#4]{%
    \def\reserved@b##1,{%
      \ifx\@nil##1\relax\else
        \ifx\relax##1\relax\else
         \noexpand\@onefilewithoptions##1[#2][#4]\noexpand\@pkgextension
        \fi
        \expandafter\reserved@b
      \fi}%
      \edef\reserved@a{\zap@space#3 \@empty}%
      \edef\reserved@a{\expandafter\reserved@b\reserved@a,\@nil,}%
  \reserved@a}

\def\zap@space#1 #2{%
  #1%
  \ifx#2\@empty\else\expandafter\zap@space\fi
  #2}

\let\@empty\empty
\def\@pkgextension{sty}

\def\@onefilewithoptions#1[#2][#3]#4{%
  \input #1.#4 }

\def\typein{%
  \let\@typein\relax
  \@testopt\@xtypein\@typein}
\def\@xtypein[#1]#2{%
  \message{#2}%
  \advance\endlinechar\@M
  \read\@inputcheck to#1%
  \advance\endlinechar-\@M
  \@typein}
\def\@namedef#1{\expandafter\def\csname #1\endcsname}
\def\@nameuse#1{\csname #1\endcsname}
\def\@cons#1#2{\begingroup\let\@elt\relax\xdef#1{#1\@elt #2}\endgroup}
\def\@car#1#2\@nil{#1}
\def\@cdr#1#2\@nil{#2}
\def\@carcube#1#2#3#4\@nil{#1#2#3}
\def\@preamblecmds{}

\def\@star@or@long#1{%
  \@ifstar
   {\let\l@ngrel@x\relax#1}%
   {\let\l@ngrel@x\long#1}}

\let\l@ngrel@x\relax
\def\newcommand{\@star@or@long\new@command}
\def\new@command#1{%
  \@testopt{\@newcommand#1}0}
\def\@newcommand#1[#2]{%
  \@ifnextchar [{\@xargdef#1[#2]}%
                {\@argdef#1[#2]}}
\long\def\@argdef#1[#2]#3{%
   \@ifdefinable #1{\@yargdef#1\@ne{#2}{#3}}}
\long\def\@xargdef#1[#2][#3]#4{%
  \@ifdefinable#1{%
     \expandafter\def\expandafter#1\expandafter{%
          \expandafter
          \@protected@testopt
          \expandafter
          #1%
          \csname\string#1\expandafter\endcsname
          {#3}}%
       \expandafter\@yargdef
          \csname\string#1\endcsname
           \tw@
           {#2}%
           {#4}}}
\def\@testopt#1#2{%
  \@ifnextchar[{#1}{#1[#2]}}
\def\@protected@testopt#1{
  \ifx\protect\@typeset@protect
    \expandafter\@testopt
  \else
    \@x@protect#1%
  \fi}
\long\def\@yargdef#1#2#3{%
  \@tempcnta#3\relax
  \advance \@tempcnta \@ne
  \let\@hash@\relax
  \edef\reserved@a{\ifx#2\tw@ [\@hash@1]\fi}%
  \@tempcntb #2%
  \@whilenum\@tempcntb <\@tempcnta
     \do{%
         \edef\reserved@a{\reserved@a\@hash@\the\@tempcntb}%
         \advance\@tempcntb \@ne}%
  \let\@hash@##%
  \l@ngrel@x\expandafter\def\expandafter#1\reserved@a}
\long\def\@reargdef#1[#2]#3{%
  \@yargdef#1\@ne{#2}{#3}}
\def\renewcommand{\@star@or@long\renew@command}
\def\renew@command#1{%
  {\escapechar\m@ne\xdef\@gtempa{{\string#1}}}%
  \expandafter\@ifundefined\@gtempa
     {\@latex@error{\string#1 undefined}\@ehc}%
     {}%
  \let\@ifdefinable\@rc@ifdefinable
  \new@command#1}
\long\def\@ifdefinable #1#2{%
      \edef\reserved@a{\expandafter\@gobble\string #1}%
     \@ifundefined\reserved@a
         {\edef\reserved@b{\expandafter\@carcube \reserved@a xxx\@nil}%
          \ifx \reserved@b\@qend \@notdefinable\else
            \ifx \reserved@a\@qrelax \@notdefinable\else
              #2%
            \fi
          \fi}%
         \@notdefinable}
\let\@@ifdefinable\@ifdefinable
\long\def\@rc@ifdefinable#1#2{%
  \let\@ifdefinable\@@ifdefinable
  #2}
\def\newenvironment{\@star@or@long\new@environment}
\def\new@environment#1{%
  \@testopt{\@newenva#1}0}
\def\@newenva#1[#2]{%
   \@ifnextchar [{\@newenvb#1[#2]}{\@newenv{#1}{[#2]}}}
\def\@newenvb#1[#2][#3]{\@newenv{#1}{[#2][#3]}}
\def\renewenvironment{\@star@or@long\renew@environment}
\def\renew@environment#1{%
  \@ifundefined{#1}%
     {\@latex@error{Environment #1 undefined}\@ehc
     }{}%
  \expandafter\let\csname#1\endcsname\relax
  \expandafter\let\csname end#1\endcsname\relax
  \new@environment{#1}}
\long\def\@newenv#1#2#3#4{%
  \@ifundefined{#1}%
    {\expandafter\let\csname#1\expandafter\endcsname
                         \csname end#1\endcsname}%
    \relax
  \expandafter\new@command
     \csname #1\endcsname#2{#3}%
     \l@ngrel@x\expandafter\def\csname end#1\endcsname{#4}}

\def\providecommand{\@star@or@long\provide@command}
\def\provide@command#1{%
  {\escapechar\m@ne\xdef\@gtempa{{\string#1}}}%
  \expandafter\@ifundefined\@gtempa
    {\def\reserved@a{\new@command#1}}%
    {\def\reserved@a{\renew@command\reserved@a}}%
   \reserved@a}%

\def\@ifundefined#1{%
  \expandafter\ifx\csname#1\endcsname\relax
    \expandafter\@firstoftwo
  \else
    \expandafter\@secondoftwo
  \fi}

\chardef\@xxxii=32
\mathchardef\@Mi=10001
\mathchardef\@Mii=10002
\mathchardef\@Miii=10003
\mathchardef\@Miv=10004

\newcount\@tempcnta
\newcount\@tempcntb
\newif\if@tempswa\@tempswatrue
\newdimen\@tempdima
\newdimen\@tempdimb
\newdimen\@tempdimc
\newbox\@tempboxa
\newskip\@tempskipa
\newskip\@tempskipb
\newtoks\@temptokena

\long\def\@whilenum#1\do #2{\ifnum #1\relax #2\relax\@iwhilenum{#1\relax
     #2\relax}\fi}
\long\def\@iwhilenum#1{\ifnum #1\expandafter\@iwhilenum
         \else\expandafter\@gobble\fi{#1}}
\long\def\@whiledim#1\do #2{\ifdim #1\relax#2\@iwhiledim{#1\relax#2}\fi}
\long\def\@iwhiledim#1{\ifdim #1\expandafter\@iwhiledim
        \else\expandafter\@gobble\fi{#1}}
\long\def\@whilesw#1\fi#2{#1#2\@iwhilesw{#1#2}\fi\fi}
\long\def\@iwhilesw#1\fi{#1\expandafter\@iwhilesw
         \else\@gobbletwo\fi{#1}\fi}
\def\@nnil{\@nil}
\def\@empty{}
\def\@fornoop#1\@@#2#3{}
\long\def\@for#1:=#2\do#3{%
  \expandafter\def\expandafter\@fortmp\expandafter{#2}%
  \ifx\@fortmp\@empty \else
    \expandafter\@forloop#2,\@nil,\@nil\@@#1{#3}\fi}
\long\def\@forloop#1,#2,#3\@@#4#5{\def#4{#1}\ifx #4\@nnil \else
       #5\def#4{#2}\ifx #4\@nnil \else#5\@iforloop #3\@@#4{#5}\fi\fi}
\long\def\@iforloop#1,#2\@@#3#4{\def#3{#1}\ifx #3\@nnil
       \expandafter\@fornoop \else
      #4\relax\expandafter\@iforloop\fi#2\@@#3{#4}}
\def\@tfor#1:={\@tf@r#1 }
\long\def\@tf@r#1#2\do#3{\def\@fortmp{#2}\ifx\@fortmp\space\else
    \@tforloop#2\@nil\@nil\@@#1{#3}\fi}
\long\def\@tforloop#1#2\@@#3#4{\def#3{#1}\ifx #3\@nnil
       \expandafter\@fornoop \else
      #4\relax\expandafter\@tforloop\fi#2\@@#3{#4}}
\long\def\@break@tfor#1\@@#2#3{\fi\fi}
\def\@removeelement#1#2#3{%
  \def\reserved@a##1,#1,##2\reserved@a{##1,##2\reserved@b}%
  \def\reserved@b##1,\reserved@b##2\reserved@b{%
    \ifx,##1\@empty\else##1\fi}%
  \edef#3{%
    \expandafter\reserved@b\reserved@a,#2,\reserved@b,#1,\reserved@a}}

\let\ExecuteOptions\@gobble

\def\@latex@error#1#2{%
  \errhelp{#2}\errmessage{#1}}

\bgroup\uccode`\!`\%\uppercase{\egroup
\def\@percentchar{!}}

\ifx\@@input\@undefined
 \let\@@input\input
\fi

\def\input{\@ifnextchar\bgroup\@iinput\@@input}
\def\@iinput#1{\input{#1} }

    \def\filename@parse#1{%
      \let\filename@area\@empty
      \expandafter\filename@simple#1.\\}

  \def\filename@simple#1.#2\\{%
    \ifx\\#2\\%
       \let\filename@ext\relax
    \else
       \edef\filename@ext{\filename@dot#2\\}%
    \fi
    \edef\filename@base{#1}}
  \def\filename@dot#1.\\{#1}

\long\def \IfFileExists#1#2#3{%
  \openin\@inputcheck#1 %
  \ifeof\@inputcheck
    \ifx\input@path\@undefined
      \def\reserved@a{#3}%
    \else
      \def\reserved@a{\@iffileonpath{#1}{#2}{#3}}%
    \fi
  \else
    \closein\@inputcheck
    \edef\@filef@und{#1 }%
    \def\reserved@a{#2}%
  \fi
  \reserved@a}
\long\def\@iffileonpath#1{%
  \let\reserved@a\@secondoftwo
  \expandafter\@tfor\expandafter\reserved@b\expandafter
             :\expandafter=\input@path\do{%
    \openin\@inputcheck\reserved@b#1 %
    \ifeof\@inputcheck\else
      \edef\@filef@und{\reserved@b#1 }%
      \let\reserved@a\@firstoftwo%
      \closein\@inputcheck
      \@break@tfor
    \fi}%
  \reserved@a}
\long\def \InputIfFileExists#1#2{%
  \IfFileExists{#1}%
    {#2\@addtofilelist{#1}\@@input \@filef@und}}

\chardef\@inputcheck0

\let\@addtofilelist \@gobble

\def\@defaultunits{\afterassignment\remove@to@nnil}
\def\remove@to@nnil#1\@nnil{}

\newdimen\leftmarginv
\newdimen\leftmarginvi

\newdimen\@ovxx
\newdimen\@ovyy
\newdimen\@ovdx
\newdimen\@ovdy
\newdimen\@ovro
\newdimen\@ovri
\newdimen\@xdim
\newdimen\@ydim
\newdimen\@linelen
\newdimen\@dashdim

\long\def\mbox#1{\leavevmode\hbox{#1}}

\let\@onlypreamble\@gobble

\let\protect\relax

\newdimen\fboxsep
\newdimen\fboxrule

\def\@height{height} \def\@depth{depth} \def\@width{width}
\def\@minus{minus}
\def\@plus{plus}
\def\hb@xt@{\hbox to}

\long\def\@begin@tempboxa#1#2{%
   \begingroup
     \setbox\@tempboxa#1{\color@begingroup#2\color@endgroup}%
     \def\width{\wd\@tempboxa}%
     \def\height{\ht\@tempboxa}%
     \def\depth{\dp\@tempboxa}%
     \let\totalheight\@ovri
     \totalheight\height
     \advance\totalheight\depth}
\let\@end@tempboxa\endgroup

\let\set@color\relax
\let\color@begingroup\relax
\let\color@endgroup\relax
\let\color@setgroup\relax

\let\color@hbox\relax
\let\color@vbox\relax
\let\color@endbox\relax

\begingroup
  \catcode`P=12
  \catcode`T=12
  \lowercase{
    \def\x{\def\rem@pt##1.##2PT{##1\ifnum##2>\z@.##2\fi}}}
  \expandafter\endgroup\x
\def\strip@pt{\expandafter\rem@pt\the}

\def\@input#1{%
  \IfFileExists{#1}{\@@input\@filef@und}{\message{No file #1.}}}

\def\@warning{\immediate\write16}

\def\Gin@driver{dvips.def}
\input graphicx.sty

\resetatcatcode

\newif\ifAMStwofonts

\ifoldfss
  \ifCUPmtlplainloaded \else
    \NewTextAlphabet{textbfit} {cmbxti10} {}
    \NewTextAlphabet{textbfss} {cmssbx10} {}
    \NewMathAlphabet{mathbfit} {cmbxti10} {} 
    \NewMathAlphabet{mathbfss} {cmssbx10} {} 
  \fi
  \ifAMStwofonts
    \ifCUPmtlplainloaded \else
      \NewSymbolFont{upmath} {eurm10}
      \NewSymbolFont{AMSa} {msam10}
      \NewMathSymbol{\upi}     {0}{upmath}{19}
      \NewMathSymbol{\umu}     {0}{upmath}{16}
      \NewMathSymbol{\upartial}{0}{upmath}{40}
      \NewMathSymbol{\leqslant}{3}{AMSa}{36}
      \NewMathSymbol{\geqslant}{3}{AMSa}{3E}

      \let\geq=\geqslant 
    \fi
  \fi
\fi 

\ifnfssone
  \newmathalphabet{\mathit}
  \addtoversion{normal}{\mathit}{cmr}{m}{it}
  \addtoversion{bold}{\mathit}{cmr}{bx}{it}
  \newmathalphabet{\mathbfit} 
  \addtoversion{normal}{\mathbfit}{cmr}{bx}{it}
  \addtoversion{bold}{\mathbfit}{cmr}{bx}{it}
  \newmathalphabet{\mathbfss} 
  \addtoversion{normal}{\mathbfss}{cmss}{bx}{n}
  \addtoversion{bold}{\mathbfss}{cmss}{bx}{n}
  \ifAMStwofonts
    \ifCUPmtlplainloaded \else
      \UseAMStwoboldmath
      \makeatletter
      \new@mathgroup\upmath@group
      \define@mathgroup\mv@normal\upmath@group{eur}{m}{n}
      \define@mathgroup\mv@bold\upmath@group{eur}{b}{n}
      \edef\UPM{\hexnumber\upmath@group}
      \new@mathgroup\amsa@group
      \define@mathgroup\mv@normal\amsa@group{msa}{m}{n}
      \define@mathgroup\mv@bold\amsa@group{msa}{m}{n}
      \edef\AMSa{\hexnumber\amsa@group}
      \makeatother
      \mathchardef\upi="0\UPM19
      \mathchardef\umu="0\UPM16
      \mathchardef\upartial="0\UPM40
      \mathchardef\leqslant="3\AMSa36
      \mathchardef\geqslant="3\AMSa3E

      \let\geq=\geqslant 
    \fi
  \fi
\fi 

\ifnfsstwo
  \DeclareMathAlphabet{\mathbfit}{OT1}{cmr}{bx}{it}
  \SetMathAlphabet\mathbfit{bold}{OT1}{cmr}{bx}{it}
  \DeclareMathAlphabet{\mathbfss}{OT1}{cmss}{bx}{n}
  \SetMathAlphabet\mathbfss{bold}{OT1}{cmss}{bx}{n}
  \ifAMStwofonts
    \ifCUPmtlplainloaded \else
      \DeclareSymbolFont{UPM}{U}{eur}{m}{n}
      \SetSymbolFont{UPM}{bold}{U}{eur}{b}{n}
      \DeclareSymbolFont{AMSa}{U}{msa}{m}{n}
      \DeclareMathSymbol{\upi}{0}{UPM}{"19}
      \DeclareMathSymbol{\umu}{0}{UPM}{"16}
      \DeclareMathSymbol{\upartial}{0}{UPM}{"40}
      \DeclareMathSymbol{\leqslant}{3}{AMSa}{"36}
      \DeclareMathSymbol{\geqslant}{3}{AMSa}{"3E}

      \let\geq=\geqslant 
    \fi
  \fi
\fi 

\ifCUPmtlplainloaded \else
  \ifAMStwofonts \else 
    \def\upi{\pi}
    \def\umu{\mu}
    \def\upartial{\partial}
  \fi
\fi

\title[Damped Lyman $\alpha$ systems and disk galaxies]{Damped Lyman $\alpha$ systems and disk galaxies: number density, column density distribution and gas density}
\author[Samuel Boissier et al.]
       {Samuel Boissier,$^1$  C\'eline P\'eroux$^{1,2}$, and Max Pettini$^1$  \\
        $^1$ Institute of Astronomy, Madingley Road, CB3 OHA, Cambridge, United Kingdom \\
	$^2$ Osservatorio Astronomico di Trieste, Via G. B. Tiepolo, 11, 34131 Trieste, Italy}

\pagerange{\pageref{firstpage}--\pageref{lastpage}}
\pubyear{2001}

\begin{document}

\def\Vc{$V_C$}
\def\l{$\lambda$}
\def\simlt{$_<\atop{^\sim}$}
\def\simgt{$_>\atop{^\sim}$}
\def\etal{{et al.\ }}
\def\Og{$\Omega_{g}$}
\def\mnras{MNRAS}
\def\apj{ApJ}
\def\apjs{ApJS}
\def\aj{AJ}

\def\ltsima{$\; \buildrel < \over \sim \;$}
\def\simlt{\lower.5ex\hbox{\ltsima}}
\def\gtsima{$\; \buildrel > \over \sim \;$}
\def\simgt{\lower.5ex\hbox{\gtsima}}
\def\arcs{$''~$}
\def\arcm{$'~$}

\maketitle

\label{firstpage}

\begin{abstract}

We present a comparison between the observed properties of damped Ly$\alpha$ systems
(DLAs) and the predictions of simple models for the evolution of present day disk galaxies, 
including both low and high surface brightness galaxies. 
We focus in particular on the number density, column density distribution and gas density
of DLAs, which have now been measured in relatively large samples of absorbers.
From the comparison we estimate the contribution of present day disk galaxies to the
population of damped Ly$\alpha$ systems, and how it varies with redshift. 
Based on the differences between the models
and the observations, we also speculate on the nature of the 
fraction of DLAs which apparently do not arise in disk galaxies.

\end{abstract}

\begin{keywords}
galaxies: evolution,
galaxies: formation,
galaxies: spiral,
quasars: absorption lines

\end{keywords}

\section{Introduction}

The true nature of the galaxies responsible for high redshift damped
Ly$\alpha$ systems (DLAs, absorbers seen in quasar spectra with
H~I column densities $N$(H~I)\,$\geq 2 \times 10^{20}$~cm$^{-2}$) is largely
unconstrained. 
At present, there are two main competing scenarios for their origin.
One school of thought sees DLAs as the (large) progenitors of
massive {\it spiral disks} (Wolfe \etal 1986; Lanzetta \etal
1991). The gas disks would have formed at $z  >5$ through monolithic
collapse, and this gas is converted to stars over a Hubble time. 
In support of this picture Prochaska \&
Wolfe (1998) argued that the kinematics of the metal absorption lines
in DLAs are best explained if they are formed in thick, large, and
rapidly rotating galactic disks, with circular velocities
$V_C  \simgt 200$\,km~s$^{-1}$.
However, this large disk hypothesis runs counter to 
currently popular models of hierarchical structure formation in which 
present day galaxies are assembled from virialized sub-units over a
protracted time interval ($z \sim 1-5$).   Haehnelt \etal (1998)
argued that hydrodynamic N-body
simulations are able to reproduce the
velocity structure of the absorption lines 
with infalling sub-galactic clumps in collapsing dark matter
haloes with small virial velocities ($V \sim 100$\,km~s$^{-1}$).

More generally, the kinematics and metallicites
of DLAs have been interpreted as evidence that they arise
in spiral galaxies (Fritze-V.Alvensleben et al. 2001; 
Hou, Boissier \& Prantzos 2001),
low surface brightness galaxies (Jimenez, Bowen \& Matteucci 1999),
dwarf galaxies (Matteucci, Molaro \& Vladilo 1997),
the progenitors of globular clusters (Burgarella, Kissler-Patig,
\& Buat 2001), the building blocks of current galaxies
(Tissera et al. 2001), galaxies undergoing tidal stripping and mergers
(Maller et al. 2001), and outflows from dwarf galaxies (Schaye 2001a).

Observationally, imaging studies of the fields of QSOs with
damped systems have shown conclusively that the absorbers
are a very `mixed bag', which includes galaxies of different
luminosities and surface brightnesses, down to objects
with apparently no associated stellar populations
which remain undetected even in very deep images
(Steidel \etal 1994, 1995; Le Brun \etal 1997; 
Lanzetta et al. 1997; Fynbo \etal 1999;  
Pettini \etal 2000; Turnshek \etal 2001;
Bowen, Tripp, \& Jenkins 2001; Kulkarni et al. 2000, 2001;
Colbert \& Malkan 2002).
While it seems clear that selection based on H~I
absorption cross-section picks out a variety of galaxies,
it is of interest to establish if one particular class of objects
dominates and, if so, whether the dominant population
changes with cosmic epoch.

In this paper we assess
the contribution of the progenitors of today's disk
galaxies to the DLA population at different redshifts
by comparing the most up to date observational determinations 
of several properties of DLAs with models
of the chemical and spectrophotometric
evolution of disk galaxies, including low surface brightness
(LSB) galaxies.
Specifically, we have compiled recent data on
the number density per unit redshift, the column density
distribution and its integral, which gives the 
total mass of H~I traced by DLA, for a large sample
of DLAs. These data are presented in \S2, while the models
of disk galaxies are discussed in \S3. In \S4
we compare our predictions with
the statistical properties of DLAs, as well as with
available imaging observations of the galaxies identified as
DLA absorbers.
We further speculate on what the 
differences we find at high redshift 
between models and observations
may be telling us about the earliest population
of DLAs. We summarise and discuss our results in \S6.

Throughout the paper we adopt the currently favoured
cosmology $H_0 = 65 $\,km~s$^{-1}$~Mpc$^{-1}$, $\Omega_M = 0.3$, 
$\Omega_{\Lambda} = 0.7$.

\section{DLA observations}
\label{secobs}

The data used in this paper are a compilation of several surveys at
high (Storrie-Lombardi \etal 1996;
Storrie-Lombardi \& Wolfe 2000; P\'eroux \etal 2002) and
low redshifts (Rao \& Turnshek 2000; Churchill
2001). From these observations we extract information on 
three properties of DLAs which our models attempt to reproduce.

The first, and most straightforward, quantity is the 
{\it number density} per unit redshift $dn/dz = n(z)$, 
plotted in Figure 1.

The second quantity is the {\it column density distribution}, 
obtained from the expression:
\begin{equation}
f(N, z) dN dX = \frac{n}{\Delta N \sum_{i=1}^{m} \Delta X_i} dN dX 
\end{equation}

\noindent where $n$ is the number of DLAs with hydrogen column density
between $N$ and $N+\Delta N$ detected in the spectra 
of $m$ QSOs encompassing a total absorption distance 
$\sum_{i=1}^{m} \Delta X_i$. The absorption distance is related 
to redshift by the expression (Peebles 1993):
\begin{equation}
X(z) = \int_0^z \frac{c}{H_0}(1+z)^2 \times E(z) dz
\end{equation}
where
\begin{equation}
E(z)=[\Omega_M (1+z)^3 + \Omega_{\Lambda}]^{-1/2}
\label{eqEz}
\end{equation}
Even with our relatively large sample, the statistics of 
$f(N, z)\,dN dX$ are still limited, and necessitate dividing
the data in only a few bins, both in $\Delta N$ and $\Delta z$
(see top panels of Figures 2 and 3). Also shown in these
panels are the fits to the column density distributions
derived by P\'eroux et al. (2002), assuming that $f(N, z)\,dN dX$
is well approximated by a $\Gamma$ function (similar to the 
galaxy luminosity function). These fits were based on  
a larger range of column densities than that shown here,
extending to the lower values of $N$(H~I) appropriate
to Lyman Limit systems 
(i.e. $N$(H~I)$ < 2 \times 10^{20}$\,cm$^{-2}$).

\begin{figure}
 \includegraphics[angle=-90,width=0.5\textwidth]{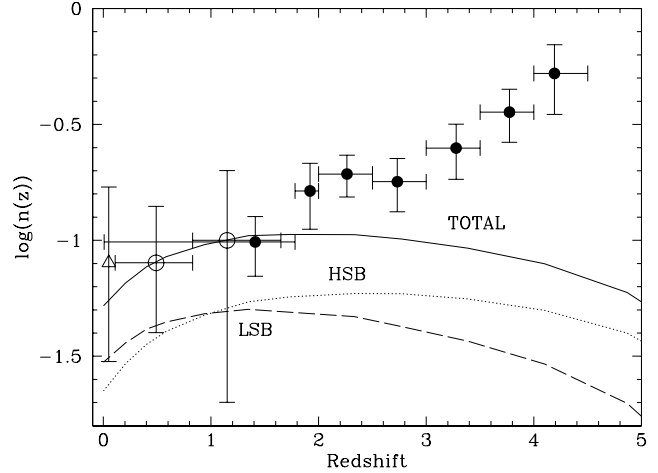}
 \caption{Redshift evolution of the number density of DLAs. The filled
 circles are data from P\'eroux et al. (2002) and references
 therein. The low redshift open circles are from Rao \& Turnshek
 (2000). The triangle at $z = 0$ is from Churchill (2001). The
 lines show the number of absorbers predicted from the models described
 in \S\ref{secmodels}. The solid line labelled `TOTAL' is 
 the prediction for all disk galaxies. The dotted and dashed lines
 show the respective contributions to this total by high and low 
 surface brightness disks. \label{fignz}}
\end{figure}

The integral of the column density distribution gives us 
the third quantity of interest, the total {\it mass
of neutral gas} traced by QSO absorbers. It is customary to
express this as a fraction of the closure density:
\begin{equation}
\Omega_{\rm DLA}(z) = \frac{ \mu m_{\rm H}}{ \rho_{crit}}
\int_{N_{min}}^{\infty} N f(N,z) dN
\label{eqn_omega}
\end{equation}
where $m_{\rm H}$ is the mass of the hydrogen atom,
$\mu$ is the mean atomic weight per baryon 
($\mu = 1.4$ for solar abundances; Anders \& Grevesse 1989)
and $\rho_{crit}$ is the closure density
($\rho_{crit} = 3 H_0^2/(8 \pi G) = 
1.96 \times 10^{-29} h^2$\,gm cm$^{-3}$,
where $h$ is the Hubble constant in units of 
100\,km~s$^{-1}$~Mpc$^{-1}$).
Values of $\Omega_{\rm DLA}$  are plotted in Figure 4.

\begin{figure}
 \includegraphics[width=0.5\textwidth]{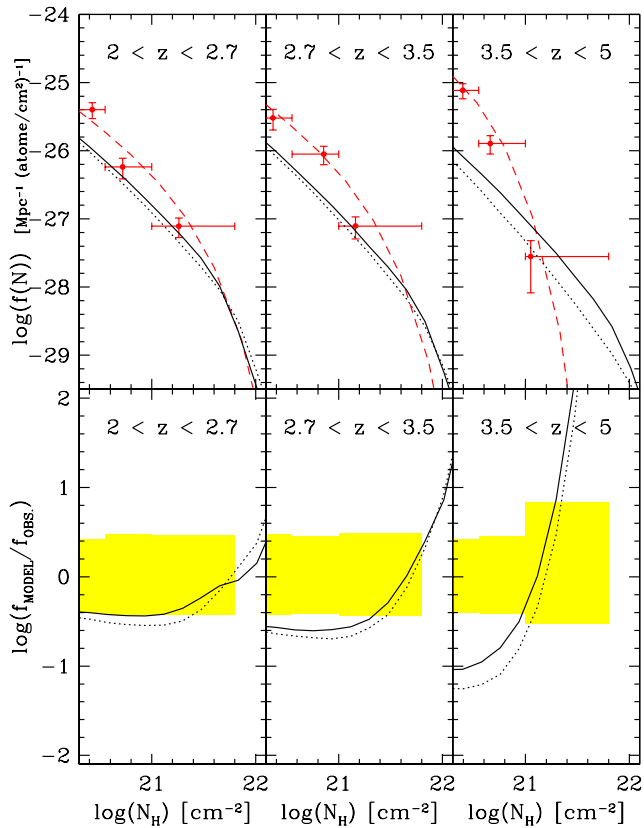}
  \caption{Column density distributions at $2<z<2.7$ (left),
  $2.7<z<3.5$ (centre) and $3.5<z<5$ (right). In the top panels, the
  points are the binned observations and the dashed lines
  shows the best fits obtained by P\'eroux et al. (2002)
  for a $\Gamma$ function extending to lower values of $N$(H~I) than those
  shown here. The solid (dotted) lines show the model predictions
  at the lower (upper) boundaries of each redshift interval
  (i.e. at $z = 2.0$ and 2.7 for the leftmost panel and so on).
  The lower panels show the differences between the model predictions
  and the $\Gamma$ function fits to the distributions;
  the shaded areas show the ranges of uncertainty arising
  from the errors in the data (as shown in the top panels)
  and the $\pm 0.3$\,dex typical uncertainty of the model predictions.}
\label{figCDDhighz}
\end{figure}

\begin{figure}
 \includegraphics[width=0.5\textwidth]{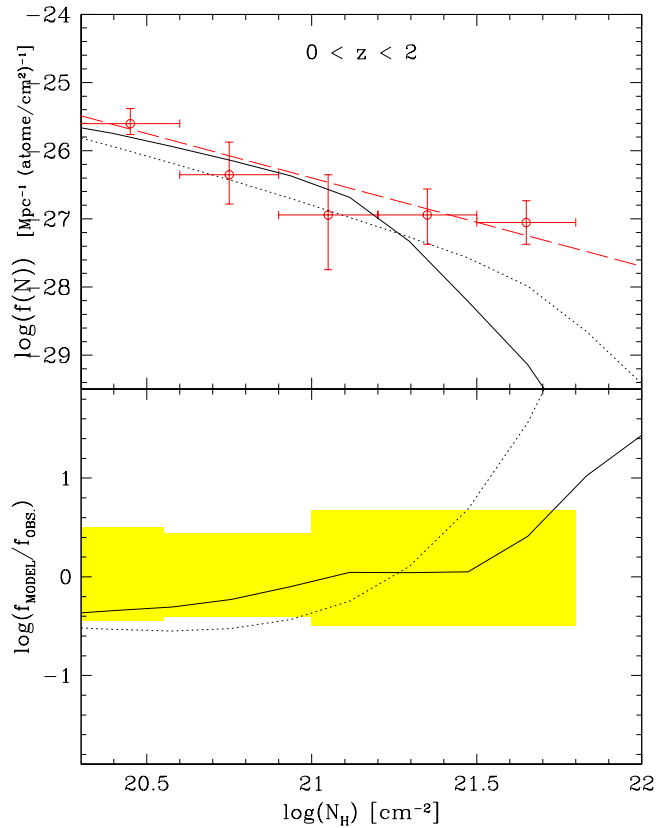}
 \caption{Column density distribution at $z<2$. As in Figure
 \ref{figCDDhighz}, the top panel shows the binned observations
 of Rao and Turnshek (2000) and the models for the redshifts
 z=0 and z=2. The dashed line is the power-law fit to the observations
 from Rao and Turnshek (2000).
 The lower panel shows the differences between the model predictions 
 at redshift 0 and 2 and the fit to the data, over the typical
 uncertainty as in Figure \ref{figCDDhighz}.
\label{figCDDlowz}}
\end{figure}

It is important to bear in mind that all the DLA surveys 
considered here used QSOs selected from magnitude limited samples.
It has long been a concern (e.g. Pei \& Fall 1995) that in these
surveys QSOs that lie behind dusty DLA systems will
be under-represented, and that
the statistics discussed above may therefore be incomplete
and biased by an unknown amount. One way to determine the
extent of such dust-induced bias is to consider the
properties of DLAs in a complete,
radio-selected, QSO sample, where dust is not an issue. 
The only such study conducted to date is the CORALS
survey by Ellison \etal (2001). Although the 
CORALS statistics are still limited,
it would appear that---at least as far as $\Omega_{\rm DLA}$
is concerned---dust bias is a relatively minor effect,
since the CORALS value is consistent with the earlier
estimates within the errors of measurement
(see lower panel of Figure 4).

\begin{figure}
 \includegraphics[width=0.5\textwidth]{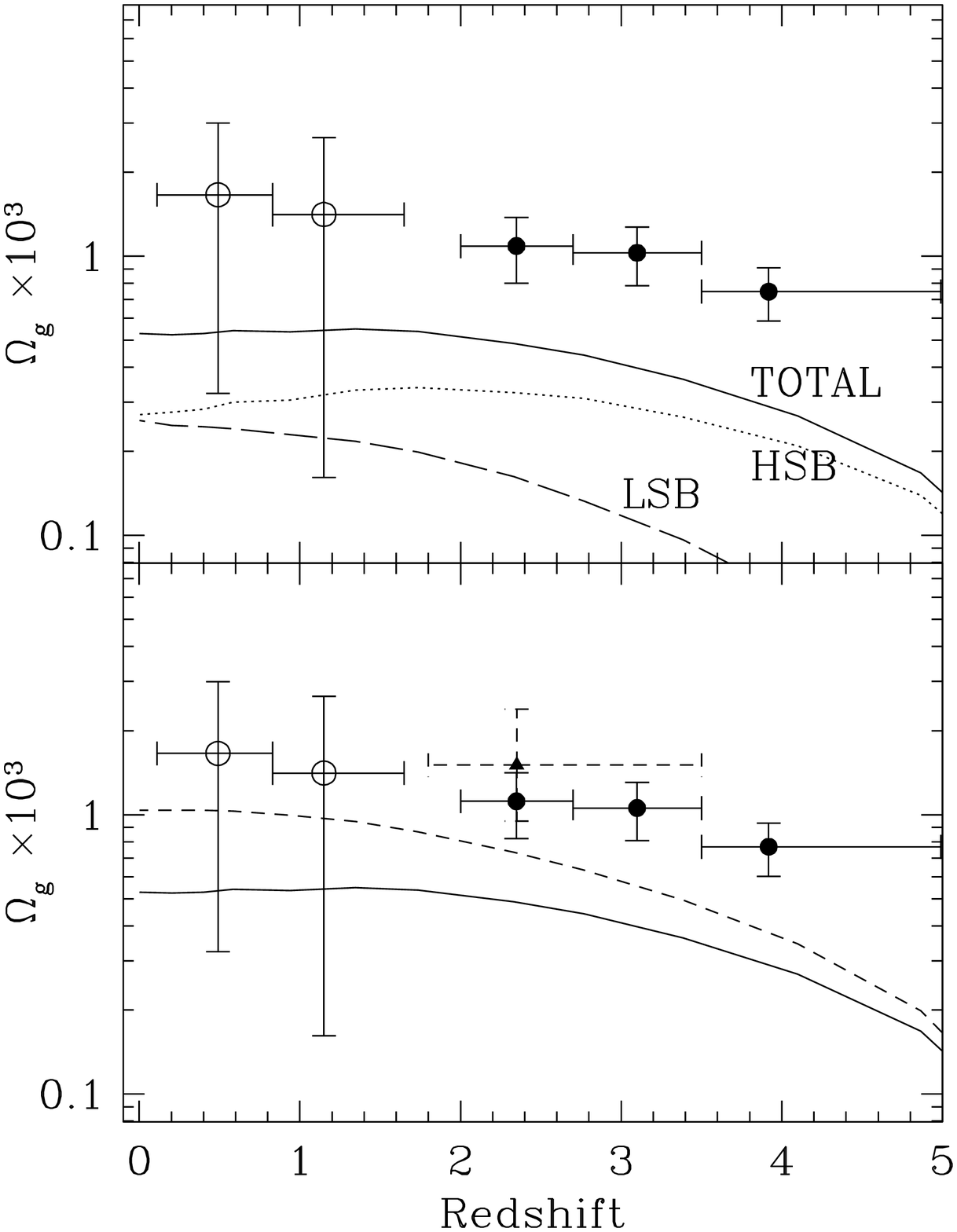}
 \caption{Redshift evolution of $\Omega_{\rm DLA}$. 
 The symbols have the same meaning as in Figure 1. 
 As in Figure 1, the lines in the top panel show our
 model predictions for all disk galaxies and separately for
 high and low surface brightness components. 
 In the lower panel, the solid line is the same
 as in the top panel, while the dashed line show the predictions
 of models without the `dust-filter' (see text). 
 The filled triangle is from the CORALS survey by 
 Ellison \etal (2001) which is free from the
 potential bias introduced by dust.} 
 \label{figomega} \label{figomegadust}
\end{figure}

\section{Models of Disk galaxies}
\label{secmodels}

\subsection{Basic Properties of the Models}
\label{secgenmod}
In Boissier \& Prantzos (2000) a `bivariate' family of 
spiral galaxies was described by considering disks with various rotational
velocities, \Vc{}, and spin parameters, \l{}.
In the local universe, the volume density $F_V$ of galaxies with a given
value of \Vc{} can be deduced from the Tully-Fisher 
relationship and the luminosity function (Gonzalez \etal 2000). 
In the following, we adopt the distribution corresponding to the parameters
listed in the fifth row of Table 4 of
Gonzalez et al., with \Vc{} between 80 and 360 km~s$^{-1}$.
The spin parameter $\lambda$ is a dimensionless measure of 
the angular momentum $J$ of the dark matter halo surrounding the disk,
and is given by:
\begin{equation}
\lambda=J |E|^{1/2} G^{-1} M^{-5/2}
\end{equation}
where $E$ and $M $ are respectively 
the energy and mass of the dark matter halo.

In the general framework of Cold Dark Matter theories,
the final properties of the disk (its mass $M_d$, scale-length $R_d$,
and central surface density $\Sigma_0$) can be derived under
simple assumptions on the relationship between the disk and 
the dark matter halo (see, for example, Mo, Mao, \&White 1998).
The scaling relations 
\begin{equation}
\frac{M_d}{M_{d,MW}}  =  \left( \frac{V_C}{V_{C,MW}} \right)^3 
\end{equation}
\begin{equation}
\frac{R_d}{R_{d,MW}}  =  \frac{V_C}{V_{C,MW}} \frac{\lambda}{\lambda_{MW}} 
\end{equation}
\begin{equation}
\frac{\Sigma_0}{\Sigma_{0,MW}} =  \frac{V_C}{V_{C,MW}} \left(\frac{\lambda}{\lambda_{MW}}\right)^{-2}
\end{equation}
(where the suffix $MW$ refers to the Milky Way) are then used
to relate these quantities to those of our Galaxy,
assumed to be a typical spiral disk.

Our models used these scaling relations to compute the chemical
evolution in zones at different distances from the centres of
disks with various rotational velocities and spin parameters.
The history of each galaxy (for a given set of values of 
$V_C$ and $\lambda$) was derived by a `backward' approach from observations of
samples of nearby galaxies and calibrated on the Milky Way 
(see Boissier \& Prantzos 1999, 2000; and Boissier \etal 2001 for more details). 
The resulting histories are characterised by an early formation of massive galaxies,
while star formation occurs on longer timescales in less massive galaxies, 
following the infall of pristine gas.

\subsection{The Spin Parameter Distribution and Low Surface Brightness Galaxies}
\label{secmodlsb}

\begin{figure}
	\includegraphics[angle=-90,width=0.5\textwidth]{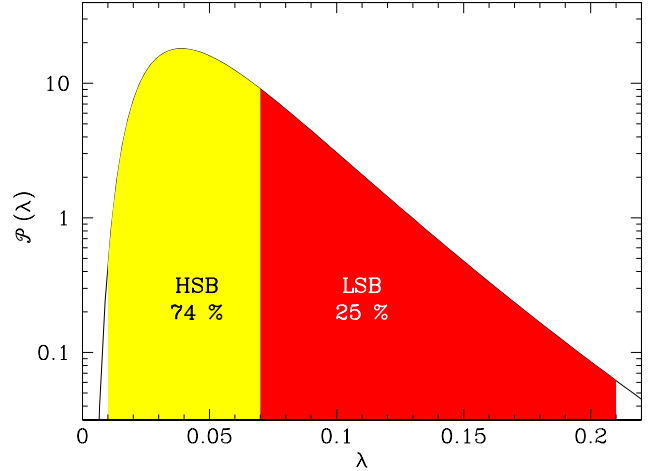}
  \caption{Distribution of the spin parameter $\lambda$. 
According to Boissier \& Prantzos (2000), galaxies with $0.01 < \lambda < 0.07$ 
are `normal' spirals, while galaxies with larger values of $\lambda$
are low surface brightness  galaxies. The respective contributions
of each (within the limits shown in the figure) to the total
integral of the distribution are indicated.
}
  \label{figspin}
\end{figure}

In our treatment we adopt the distribution of spin parameters
derived from N-body simulations (Mo et al. 1998) and reproduced
in Figure 5.
Boissier \& Prantzos (2000) considered values for the spin parameter
between $\lambda = 0.01$ and 0.09 and found that, within
the framework of simple models of chemical and spectrophotometric evolution
of spirals, values of $\lambda > 0.07$ apply to low surface brightness
galaxies, with blue central surface brightness greater than 22.5 mag arcsec$^{-2}$,
while galaxies with $\lambda < 0.07$ are `normal' spirals.
As can be seen from Figure 5, the former account for 
25\% of the total number of galaxies in the distribution
(integrating between $0.07 < \lambda < 0.21$),
while the latter contribute 74\% 
(integrating between $0.01 < \lambda < 0.07$).
In this work we assumed that this population of LSB disks does
indeed exist and we accordingly extended the models of
Boissier et al. (2001) to larger values of \l{} than
those considered in that paper in order to take fully
into account this LSB population.
Furthermore, we renormalised the distribution
in Figure 5 so that 
its integral over the range of values of $\lambda$ appropriate to
normal, high surface brightness, disks ($0.01 < \lambda < 0.07$)
is equal to 1. This is because 
the normalisation of the distribution of velocities $F_V$, 
derived from the observed galaxy luminosity function, 
does not include low surface brightness galaxies.
While this treatment of LSB galaxies is simplistic,
it does nevertheless allow us to test, in the framework of our
model, their contribution to the DLA absorber population.

\subsection{Model Predictions for the Number Density, Column Density Distribution,
and Total Gas Mass of DLAs}

\label{secmodelpredict}

\begin{figure*}
        \vspace{-3cm}
	\hspace{-1cm}
        \centering
        \includegraphics[angle=0,width=1.0\textwidth]{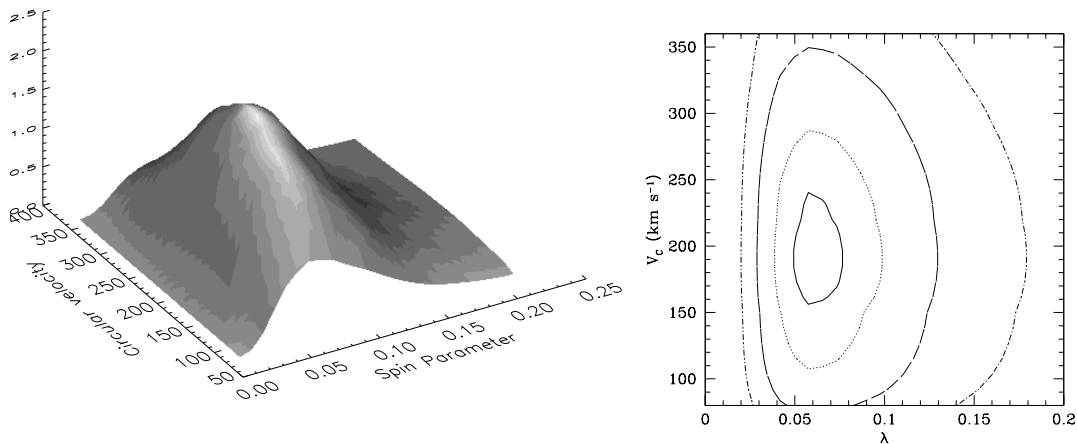}
        \vspace{-15cm}  
\caption{Surface and contour plots of the galactic cross-section 
(independent of how the mass is distributed between stars and interstellar gas) 
as a function of spin parameter $\lambda$ and 
rotational velocity $V_C$.}
  \label{figbivariate}
\end{figure*}

Prantzos \& Boissier (2000) and Hou \etal (2001) 
considered the chemical evolution of 
the family of models of disk galaxies
described above. In the zones within the galaxies
where the gas density is sufficiently high
to give a column density of neutral gas
$N$(H~I) $>$ 2 $\times$ 10$^{20}$ cm$^{-2}$, 
the models match the chemical properties of DLAs,
both in the overall degree of metal enrichment and in the
more detailed abundance pattern of different chemical elements.
However, in those treatments good agreement could only 
be achieved by imposing a `dust filter', 
whereby DLAs with large column densities
of metals (and therefore presumably dust) are excluded 
from the samples, for the reasons explained in \S2.
This dust filter was based on the analysis by 
Boiss\'e et al. (1998) and takes the form
\begin{equation}
\label{eqcondi2}
\log{\rm (}N{\rm (HI)}+{\rm [Zn/H]}) < 21
\end{equation}
where [Zn/H] is the abundance of Zn relative
to solar (in the usual notation)
and $N{\rm (HI)}$ is the 
column density of hydrogen atoms (in units cm$^{-2}$). 
The filter operates in such a way that zones within
a model galaxy which do not satisfy eq. (9)
were not taken into account, on the assumption that
such DLAs have been overlooked in existing surveys.

Here we extend the work of Prantzos \& Boissier (2000)
and Hou et al. (2001) by using the models to make
predictions for other properties of DLAs (specifically
those discussed in \S2), with the aim of assessing 
what fraction of DLAs are the precursors of today's 
disk galaxies. We adopt the same `backward' approach
as in the earlier work and, like those authors,
consider the effect of a dust filter defined as in 
eq. (9). Our treatment is broadly similar 
to that first adopted by Lanzetta (1993), but with the
following major improvements: (i) we use models which predict
the evolution of present day galaxies back in time;
(ii) we include in the models of a variety of
galaxy morphologies, masses, sizes, and surface brighteness;
and (iii) we are able to compare with the much larger data 
base of measurements of DLAs available now, 
ten years since the pioneering work of Lanzetta (1993).

The number density of DLAs per unit redshift can be expressed as (Peebles 1993)
\begin{equation}
n(z) = \frac{dN}{dz} = \left ( \frac{dX}{dz} \times \frac{dN}{dX} \right ) 
      =   \frac{c}{H_0}(1+z)^2 \times E(z) \times G(z) 
\label{nzequa}
\end{equation}
with $E(z)$ as defined in eq.(\ref{eqEz}).\footnote{
Note that the predicted value of $\frac{dN}{dz}$ therefore depends on
the cosmological model assumed, whereas the observed value does not;
the opposite is true for the column density distribution
and $\Omega_{\rm DLA}$.} 
The quantity $G(z)$ is the product of the 
volume density of galaxies
and their cross-sections, averaged 
over the whole population 
\begin{equation}
\label{eqG}
G(z)= \int_{\lambda} \int_{V_C} d^2G_{(\lambda,V)}
\end{equation}
where $d^2G_{(\lambda,V)}$ is the
contribution to the total cross-section by galaxies with 
spin parameter $\lambda$ and rotational velocity $V_C$,
given by
\begin{equation}
\label{eqcross}
d^2G_{(\lambda,V_C)} =F_V(V_C) dV F_{\lambda}(\lambda) d\lambda
\int_{S} \int_{i=0}^{i=\pi/2}  \Phi \, dS \, \frac{2}{\pi}di 
\end{equation}
Here $\int_{S} dS$ represents the integral over the apparent surface of the galaxy,
and $\int_{i} di$ is the integration over all inclination angles.
The parameter $\Phi$ is a switch which can take the values of 1 or 0.
For example, it can be set to 1 independently of all other parameters
in order to compute the geometrical 
cross section. The value of $\Phi$ (1 or 0) can also 
be determined by the inclination and the local properties
of galaxies (metallicity and gas density, both functions of the
distance from the galactic center)
to exclude zones in galaxies which
do not meet the DLA selection criteria, 
because either of the conditions
$N$(H~I)$ > 2 \times 10^{20}$\,cm$^{-2}$ or 
$\log{\rm (}N{\rm (HI)}+{\rm [Zn/H]}) < 21$
(the dust filter in eq. 9)
is not satisfied. Then, $\Phi$ is set to 0 in these zones, 
and 1 elsewhere.

Figure \ref{figbivariate} shows the function $d^2G_{(\lambda,V_C)}$
for $\Phi=1$. From this figure it can be seen that,
in our models, the major contribution to the cross-section for absorption 
(taken at this stage to be five times the scale length of the
radial distribution of mass without considering separately
the distributions of the stellar and gaseous components)
is from galaxies with relatively high rotational velocities
($V_C \simeq 200$\,km~s$^{-1}$) and intermediate values
of spin parameter ($\lambda \simeq 0.07$),
at the border line between high and low surface brightness galaxies.
This is straightforward to understand, since it is the product of
$V_C$ and $\lambda$ which determines the linear dimensions of 
galaxies. From the figure it would also appear that the ranges of values of 
$V_C$ and $\lambda$ adopted are sufficiently broad
to account for most of the cross-section. However, we may have underestimated
the total cross-section if the faint end slope of the luminosity
function is sufficiently steep that dwarf galaxies
with $V_C < 80$\,km~s$^{-1}$ make a non-negligible contribution.
Such galaxies are not included in the models considered here because
we are chiefly concerned with disk galaxies.

We predict the column density distribution numerically by
computing for each interval of column density [$N_H$, $N_H+\delta N_H$] 
the number of systems given by 
eqs. (\ref{eqG}) and (\ref{eqcross}).
The total neutral gas mass in DLAs is calculated from the expression
\begin{equation}
\Omega_G=\frac{1}{\rho_{crit}} \int_{\lambda,V} F_V(V_C) dV F_{\lambda}(\lambda)d\lambda \int_{S,i} \Phi \Sigma_G dS \frac{2}{\pi} di,
\label{eqomegatheo}
\end{equation}
which is similar to eqs. (\ref{eqG}) and (\ref{eqcross}), but with the
second integral now inluding  $\Sigma_G$, the mass surface density
along the line of sight. 
The result of eq. (\ref{eqomegatheo}) can then be compared with
the observed value of $\Omega_{\rm DLA}$, as given by 
eq. (\ref{eqn_omega}).
In both cases, $\Phi$ was set to 0 if 
$N$(H~I)\,$ < 2 \times 10^{20}$\,cm$^{-2}$ or if
the condition imposed by eq. (9) was not met.

\subsection{Neutral Atomic Gas}

The models of chemical evolution which we use to describe disk
galaxies compute the surface density of the {\it total} amount of 
interstellar matter, in all its forms. On the other hand,
for the comparison with the observed properties of DLAs
we are interested only in the neutral atomic gas component. 
We do not consider the fraction of gas in ionised form
because it is likely to be unimportant for our purposes.
In the Galactic disk, for example, ionised gas accounts for less
than 1\% of the total gas mass of $3 \times 10^{9}\,M_{\odot}$
(Osterbrock 1974).
Similarly, we can ignore the mass fraction in solid form
(dust grains), which for solar composition is less than
2\% (Anders \& Grevesse 1989) and would be even lower at
the low metallicities typical of DLAs (e.g. Pettini et al. 1997).

The correction for molecular gas is also likely to be 
of minor importance for the following reasons.
From straightforward geometrical considerations
the cross-section for absorption is always dominated
by the outer regions of the disks where, in local galaxies
at least, the molecular fraction is low. 
In the Milky Way, for example, H$_2$ makes a significant contribution to
the total gas profile only at small radii (in a molecular ring
at $\sim 4$\,kpc), while the outer disk is totally dominated by 
atomic gas (e.g. Dame 1993). 
CO observations of nearby spiral galaxies 
show that the molecular gas density decreases more steeply
with radius than the atomic gas, and becomes a minor component beyond
a few kpc (e.g. Boissier \etal 2002). In LSB galaxies the molecular fraction
is observed to be very small (e.g. Matthews \& Gao 2001).
In any case, it is now well established that 
in DLAs $N$(H$_2$)\,$\ll N$(H~I) (Petitjean, Srianand \& Ledoux 2002).
Thus, in our models we take $N$(H~I)\, = $N$(H$_{\rm TOT}$)
(but correct the total gas density for the helium fraction).
When we tried to compute the molecular fraction,
using a recipe which links  H$_2$ to the star formation 
and calibrating on the Milky Way, 
we obtained results which are indistinguishable 
from those obtained assuming that all the gas is in atomic form.

\section{The Contribution of Disk Galaxies to the Population of DLAs}

\label{seccompa}

We are now ready to compare the output of our models with the
statistical properties of DLAs summarised in \S2. The results
are illustrated in Figures 1--4 and are discussed individually 
below.

\subsection{Number Density}

As can be seen from Figure \ref{fignz}, our models 
reproduce satisfactorily (bearing in mind the large
uncertainties in the data) the observed number density of DLAs 
per unit redshift in the interval $0 < z < 2$.
It is interesting to note that our simplistic
treatment of LSB galaxies, which essentially extrapolates
the properties of normal spirals to larger
values of the spin parameter $\lambda$ (see 
\S3.2), predicts approximately equal contributions
to $n$($z$) from high and low surface brighteness
galaxies over this redshift interval. 
As we shall see, the 
available imaging data
summarised in \S5.1  are consistent
with this rough breakdown.

Figure \ref{fignz} also shows that our models 
{\it under}predict  $n$($z$) at $z \simgt 2$
by factors of several; while the data show 
a continuous increase with look-back time,
the predicted number density flattens off and decreases
beyond $z = 2$. In reality, the observed increase
in $n$($z$) with $z$ is consistent with the
no-evolution scenario where the product of the 
comoving number density and cross-section of DLA
absorbers remains constant in time---the measured $n$($z$)
increases simply because the scale factor
of the universe decreases with $z$. Thus, the
flattening of $n$($z$) in the calculations reflects
the delayed and protracted epoch of disk formation
in the models. From this we conclude that
either today's disk galaxies were assembled
earlier than predicted by our models,
or a different class of objects is responsible for 
the majority of damped systems observed at $z > 2$.

\subsection{Column Density Distribution}

The observed and predicted column density distributions
are compared in Figures \ref{figCDDhighz} and \ref{figCDDlowz}.
At $z =  2 - 3$  the agreement
is fair, considering the large uncertainties
in the data. The models tend to predict more
high $N$(H~I) DLAs than observed, but this could
be due to small number statistics
(although it is interesting that a similar discrepancy
is found with N-body simulations---see 
Katz et al. 1996; Schaye 2001b; and Cen et al. 2002
for relevant discussions). 
At $z < 2$ the agreement is also acceptable (Figure \ref{figCDDlowz}).
Again we find a discrepancy at the high column density
end of the distribution, but this time 
the models tend to underpredict
the number of DLAs with log\,$N$(H~I)\,$\simgt 21.3$ [$N$(H~I) in cm$^{-2}$]
compared with the observations by Rao \& Turnshek (2000).
It would clearly be highly desirable to improve the 
statistics at this end of the distribution 
with more extensive surveys for DLAs at all redshits.

Returning to Figure \ref{figCDDhighz}, 
it appears that the agreement between the models and 
the data becomes progressively worse as we move to $z \simgt 3$;
the steepening of $f$($N$) with increasing $z$ uncovered
by P\'eroux et al. (2002) is not reproduced by the models.
Taken at face value, this is another indication of a change
in the population of galaxies responsible for
DLAs; while the precursors of today's disks
can account for the column density distribution
at low and intermediate redshifts, at $z \simgt 3$
an additional population, with typically lower values
of $N$(H~I), seems to take over.

\subsection{Mass Density}

From Figure \ref{figomega}
it can be seen that our disk models
underpredict $\Omega_{\rm DLA}$ 
by about a factor of two at all redshifts,
although at $z \simlt 2$ the 
uncertainties in the determination of this quantity
are so large that they do encompass our model
predictions.

All of the above results were obtained with the dust filter
described in \S3.3 (eq. 9). The lower panel of Figure \ref{figomega}
shows the effect of removing such a filter and including in the
statistics {\it all} DLAs, irrespectively of their metallicity and column 
density. We find that although $\Omega_{\rm DLA}$ does increase,
the effect is not large, amounting to an upwards
correction by a factor of $\simlt 2$ at $z = 2 - 3$,
in good agreement with
the empirical finding by Ellison et al. (2001).
In retrospect this is not surprising. The condition 
imposed by eq. (9), 
log($N$(H~I))+[Zn/H])\,$ < 21$, is {\it not}
satisfied  primarily in 
massive galaxies (statistically less favoured by
the shape of the luminosity function) 
and in the inner regions of galaxies
(statistically less favoured because they present 
a smaller cross-section than the outer regions).
It is thus quite
natural that the dust filter should {\it not} remove
a major fraction of potential DLAs.
Figure \ref{figomega} shows the dust bias being most
important at $z < 2$. It remains to be seen, 
once the statistics of DLAs at these redshifts
are improved, whether this is indeed the case.
At present the uncertainties in the 
values of $\Omega_{\rm DLA}$ deduced by Rao \& Turnshek (2000)
are so large that either set of models (with and without
dust obscuration) is consistent with the observations.

\subsection{Comparison with the Earlier Analysis by Lanzetta}

\begin{figure}
	\includegraphics[width=0.5\textwidth]{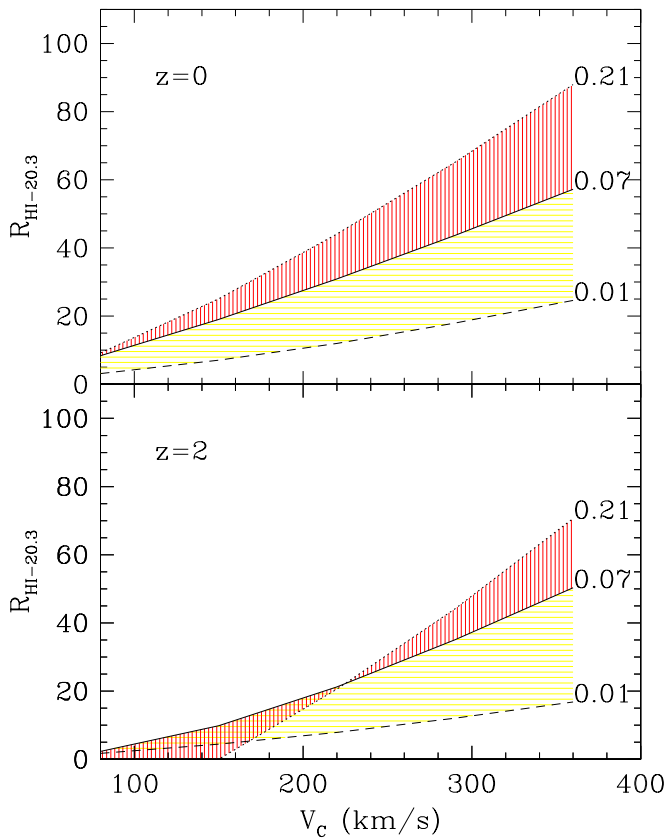}
    \caption{H~I sizes (radii at which 
$N$(H~I)\,$ = 2 \times 10^{20}$\,cm$^{-2}$)
for face-on galaxies as a function of rotational velocity
$V_C$ for three different values of the spin parameter,
 $\lambda = 0.07$ (solid line, considered here as the transition value between
 HSB and LSB spirals), 0.21 (dotted) and
  0.01. (dashed). `Normal' spirals occupy the horizontally hatched areas of the plot,
while LSB galaxies fall in the vertically hatched regions).}

  \label{figrhi}
\end{figure}

As mentioned above, our treatment is similar to that first
originally implemented by Lanzetta (1993), although we have
introduced several important refinements. It is thus of interest
to compare our conclusions with his.
In particular, Lanzetta was forced to conclude that the incidence
of DLAs required gaseous disks with typical radii of $\sim 30$\,kpc,
greater than the typical dimensions of galaxies today, both
in their stellar Holmberg radii and in their cross-sections
at $N$(H~I)\,$= 2 \times 10^{20}$\,cm$^{-2}$.
Figure \ref{figrhi} shows the cross-sections of galaxies in
our models as a function of rotational velocity and spin parameter,
at two different redshifts. Considering these plots in conjunction
with those in Figure \ref{figbivariate}, it can be realised
that such large radii are in fact not unexpected in our models
once the contribution by LSB galaxies to the DLA population
is taken into account.

\subsection{Comparison with Imaging Data at Low and Intermediate Redshifts}
\label{secspeculate}

\begin{table*}
\begin{tabular} { l l l l l l l l l }
\hline
QSO             &  $z$      & $D$ (kpc) & $M_B$ & $M_K$       &$L/L_*$  & Type    & Confirmed & Refs \\    
\hline
HS1543+5921     & 0.009     & 0.5     & $-15.7$ & $-19.3$   & 0.01  & LSB           & yes  & (a) \\
OI363           & 0.0912    & $< 3.7$ &         & $-22.1$   & 0.12  & LSB           &      & (b)  \\
OI363           & 0.2213    & 23      &         & $-22.4$   & 0.16  & Dwarf         & yes  & (b)  \\
0952+179        & 0.239     & 8.1     &         & $<-20.1$ & $<0.02$  & Patchy/irr LSB&      & (b,g) \\
1127$-$145      & 0.313     & 19.7    &         & $-20.9$   & 0.04  & Patchy/irr LSB& yes  & (b,g) \\
PKS1229$-$021   & 0.3949    & 8       & $-19.8$ &           & 0.6   & Irr.          &      & (c,d) \\
3C196           & 0.437     & 9       & $-21.5$ &           & 2.8  & Spiral        &      & (c) \\
0827+243        & 0.518     & 40.2    & $-21.3$ & $-25.3$   & 2.3   & Spiral?       & yes  & (b,g) \\
Q1629+120       & 0.532     & 20.3    & $-20.8$ & $-24.8$   & 1.5   & Spiral        &      &(g) \\
0058+019        & 0.6118    & 8.7     & $-19.8$ &           & 0.6   & Spiral        &      &(e) \\
Q1209+107       & 0.6295    & 11.9    & $-21.1$ &           & 1.9   & Spiral        &      & (c) \\
3C286           & 0.692     & 19.1    & $-20.7$ &           & 1.3   & LSB           &      & (d) \\
MC1331+170      & 0.7443    & 30.4    & $-21.6$ &           & 3.0   & Edge-on Spiral&      & (c) \\
PKS0454+039     & 0.8596    & 6.6     & $-19.9$ &           & 0.6   & Compact       &      & (c) \\
\hline
\end{tabular}
\caption{\label{tableimage} DLAs and candidate galaxies. 
References:
(a) Bowen \etal (2001); 
(b) Turnshek \etal (2001); 
(c) Le Brun \etal (1997);
(d) Steidel \etal (1994);
(e) Pettini \etal (2000);
(f) Steidel \etal (1997);
(g) Nestor \etal (2001).
}
\end{table*}

In the last few years there have been many attempts
to identify the galaxies responsible for producing
damped Ly$\alpha$ systems by deep imaging.
We have collected in Table 1 all available data
from the literature pertaining to DLAs at $z_{\rm abs} < 1$
(where the results are most straightforward to interpret),
restricting ourselves to cases where only one galaxy 
was found close to the QSO sight-line.
Column (3) gives the projected QSO-galaxy separations
in our $H_0 = 65 $\,km~s$^{-1}$~Mpc$^{-1}$, $\Omega_M = 0.3$, 
$\Omega_{\Lambda} = 0.7$ cosmology.
Column (4) and (5) list the absolute magnitudes of the putative
absorbers in this cosmology, in either the rest frame
$B$ or $K$ band; in deriving these values from the 
observed magnitudes we applied $K$-corrections 
appropriate to the SED of an Sbc 
galaxy (Bruzual \& Charlot 1993).
The galaxy luminosities are given in units of $L^{\ast}$
in column (6), adopting $M^{\ast}_{\rm K} - 5\,{\rm log}\,h = -23.44$
(Cole et al. 2001) and $M^{\ast}_{\rm B} - 5\,{\rm log}\,h = -19.46$
(Norberg et al. 2002). Column (7) gives an indication
of the galaxy morphology as reported in the original sources
referenced in the last column of the Table; cases where 
the identity of the absorber has been confirmed with 
follow-up spectroscopy are indicated in column (8).

The data collected in Table 1 form a small and far from homogeneous
sample. Furthermore, in most cases there
is no spectroscopic confirmation that the closest galaxy
to the QSO sight-line is indeed the DLA absorber.
Nevertheless, these observations are all we have at our
disposal at present, and it is still worthwhile to 
consider qualitatively how they compare with the
predictions of our models. Considering first the
galaxy morphologies, the sample includes six 
apparently `normal' spirals and five LSBs. This breakdown
is broadly in line with our expectations from
Figure 1 at $z < 1$.\footnote{It should be noted here that
the definition of LSB in our models is based 
on whether the spin parameter $\lambda$ is larger 
than 0.07\,. Owing to the 
larger star formation rate in the past in the models,
some galaxies with $\lambda > 0.07$ at $z \simeq 1$
actually have values of central surface brightness 
somewhat greater than 22.5\,mag\,arcsec$^{-2}$. However,
this is not expected to affect significantly the 
finding that spirals and LSB galaxies 
make comparable contributions to the DLA cross-section
because the morphology classifications listed in Table 1 
are only approximate estimates.}
A further three galaxies 
do not appear to be spiral disks but have been 
classified by the original observers as
irregular/dwarf/compact. The fraction may be higher because
our sample in Table 1 does not include non-detections;
nevertheless, preliminary indications are that this 
additional population of galaxies---which
we have not considered here---is unlikely to make
a major contribution to the DLA statistics.
Only a uniform survey to well defined detection limits
can address this question quantitatively.

\begin{figure}
 \includegraphics[width=0.5\textwidth]{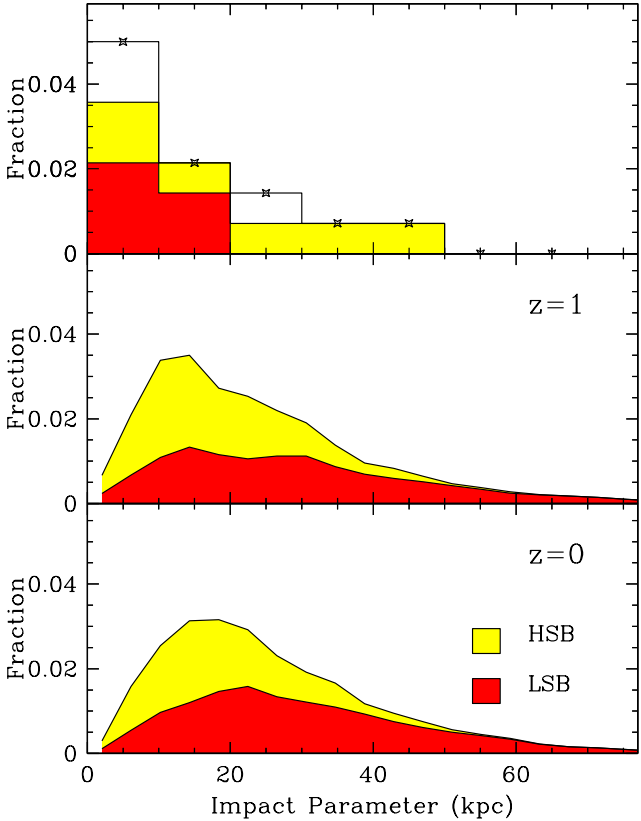}
  \caption{Distributions of impact parameters. Top panel: distribution
  constructed from Table \ref{tableimage} grouping the data 
  in 10\,kpc bins. The white areas of the histogram correspond
  to galaxies classified as irregular/dwarf/compact.  Middle and
  bottom panels: distributions predicted by our disk models at
  $z = 1$ and 0.} \label{figimpact}
\end{figure}

Figure \ref{figimpact} compares the observed and predicted
distributions of projected impact parameters $D$; the latter were calculated
for random orientations of the galaxies on the plane of the sky.
Within the limitations imposed by the small size
of the sample, the observations apparently favour
smaller values of $D$ than predicted. In some cases,
misidentifications
of the closest galaxy to the QSO sight-line as the absorber
may be the reason for this apparent discrepancy. 
The models predict an extended tail at
large impact parameters which cannot yet be discerned with
the limited data available; it will be of interest to
test this prediction with future surveys.

\begin{figure}
 \includegraphics[width=0.5\textwidth]{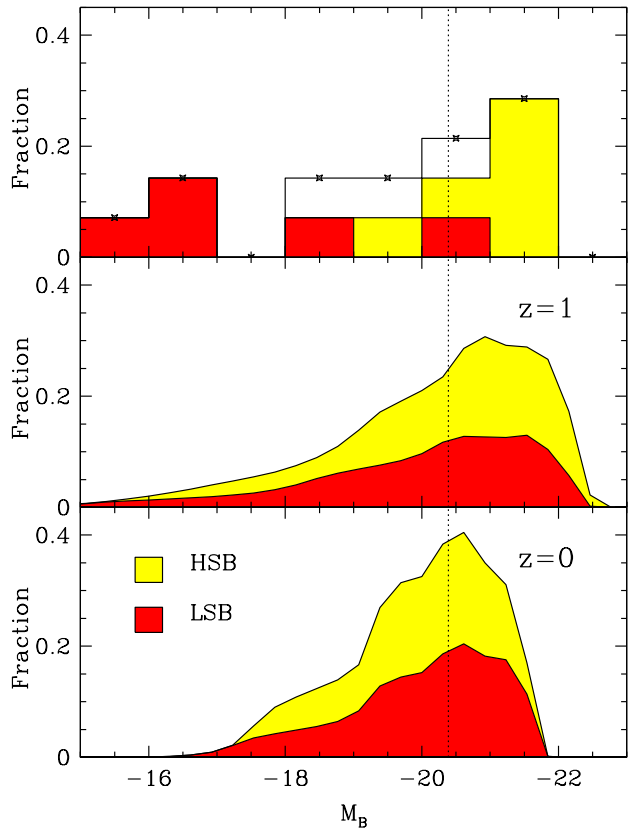}
  \caption{Distributions of absolute $B$-band magnitudes. 
  Top panel: distribution
  constructed from Table \ref{tableimage} grouping the data 
  in 1\,magnitude bins. The white areas of the histogram correspond
  to galaxies classified as irregular/dwarf/compact.  Middle and
  bottom panels: distributions predicted by our disk models at
  $z = 1$ and 0. The vertical dotted line at $M^{\ast}_{\rm B} = -20.4$
  indicates the value of $M^{\ast}_{\rm B}$ in
  our cosmology.} \label{figlumin}
\end{figure}

Figure \ref{figlumin} compares the observed and predicted
distributions of $M_B$. To improve the limited statistics,
we have used the values of $L/L^*$ in column (6) of Table 1
(which in all but three cases were based on 
{\it either} $B$-band {\it or} $K$-band
magnitudes, but did agree in the three cases
where magnitudes in both bands are available)
to calculate $M_B$ from the relation
$M^{\ast}_{\rm B} - 5\,{\rm log}\,h = -19.46$
(Norberg et al. 2002).
Evidently, our models
of disk galaxies can reproduce reasonably well the
magnitudes of the brighter galaxies associated with DLAs,
but do not extend to the low luminosities of the faintest
absorbers in the sample.  
Again, the relatively high frequency of faint LBGs
in the present sample of DLA galaxies
needs to be investigated further with more extensive
imaging surveys before firm conclusions can be drawn.

\section{The Population of DLAs at High Redshifts}

\begin{figure}
 \includegraphics[width=0.5\textwidth]{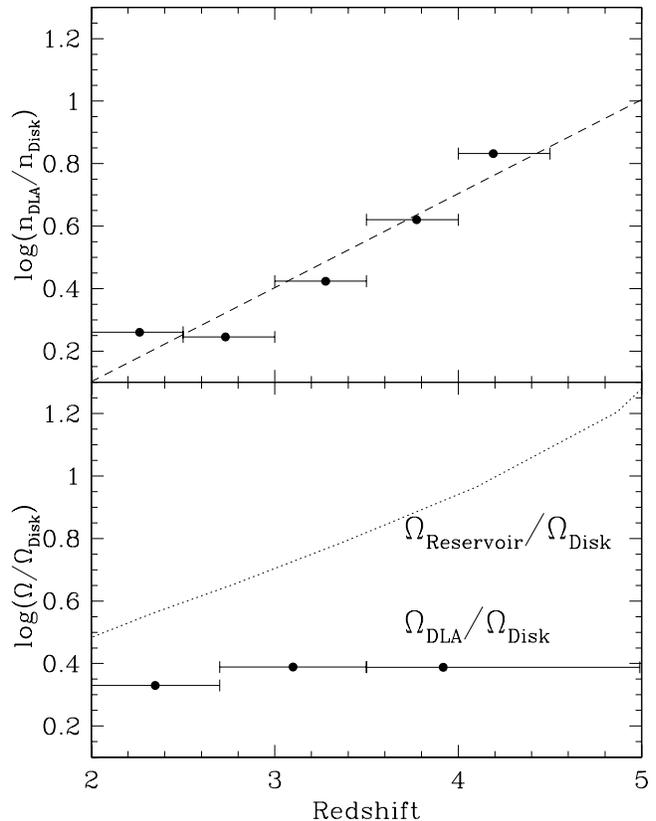}
  \caption{{\it Top:\/} The ratio between the observed number density of
DLAs per unit redshift and the value predicted by our disk galaxy models.
The data suggest the existence of a rapidly evolving `excess' population
of DLAs at high $z$ over what can be accounted for by extrapolating
the properties of today's spiral galaxies back in time.
The dashed line corresponds to a halving of $n$($z$) for this additional
population per unit redshift between $z = 4.5$ and 2.
{\it Bottom:\/} Ratio between the mass density of the total gas
reservoir available for disk formation and that which has 
been incorporated into disks (dashed line); and between the 
observed mass density in DLAs and that in our model disks
(filled circles).} \label{fighigzdif}
\end{figure}

As discussed in \S4 and shown in Figures 1--4, our models
for disk galaxies provide a progressively poorer match
to observations of DLAs beyond $z \simeq 2 - 3$.
Taken at face value, this may indicate that the precursors
of present day spiral galaxies---evolved back in time according to
our models which do not include merging---are insufficient to
account for the observed numbers of DLAs and that an additional
population of absorbers is required.
We investigate this point further in Figure \ref{fighigzdif}.
In the top panel we show, on a log scale and as a function
of redshift, the ratio of the observed number of 
DLAs (per unit redshift) to the number which can be attributed
to the progenitors of today's disks according to
our models. Thus this plot shows the ratios between the
data points and the line labelled `TOTAL' in Figure 1. 
While at $z \simeq 2$ disks can account for the number
density of DLAs within the uncertainties in both the data 
and the model predictions, at redshift $z \simeq 4.5$
DLAs are nearly ten times more numerous than predicted.
This excess evidently decreases rapidly with time,
approximately by a factor of two per unit redshift 
in the interval $2 < z < 4.5$ (dashed line).

The bottom panel of Figure \ref{fighigzdif}
shows the contribution of our model disk galaxies
to the observed value of $\Omega_{\rm DLA}$
(the ratios between the data points and the 
line labelled `TOTAL' in the top panel of 
Figure 4). 
Also shown in this panel is the progressive
transfer of baryons from the overall
gas reservoir---from which disks are formed
in our models---into galaxies, as disks
form and grow with time. Evidently,
disks can account for approximately
half of $\Omega_{\rm DLA}$ at all epochs
(as already shown in Figure 4).
In other words, it appears that the additional 
population of absorbers which dominate the number counts 
of DLAs at the highest redshifts do not make a significant
contribution to $\Omega_{\rm DLA}$ compared with
the baryons which are in disks. This is reflected by
the steepening of the column density distribution
with increasing redshift found by  P\'eroux et al. (2002)
and shown in Figure 2.

A possible interpretation of these results is that the 
`excess' population of DLAs at high redshift are 
small structures, of relatively low column density,
which progressively merge with each other and with more massive
galaxies to form the progenitors of today's disk galaxies
by $z \simeq 2$. This scenario, which at the moment can 
only be regarded as speculative, would explain at the same
time the observed decrease of $n$($z$) and 
flattening of the column density distribution (resulting
in relatively more numerous systems of high column density) 
with the progress of time from $z = 4.5$ to 2. Observationally,
this scenario is consistent with:
(a) the initial results
of imaging searches for the galaxies responsible for DLAs
at $z \simeq 4$ which suggest that they are fainter than
$\sim 1/4 L^{\ast}$ (Prochaska et al. 2002); and
(b) the low metallicity---typically 1/100 of solar, only marginally
higher than that of the intergalactic medium 
traced by the Ly$\alpha$ forest---of DLAs
at these high redshifts (Prochaska \& Wolfe 2002).

\section{Summary and Conclusions}
\label{summary}

In this paper we have used the models developed by
Boissier \& Prantzos (2000), which describe the 
chemical and spectrophotometric evolution of disk galaxies
of different surface brightnesses, to assess whether
the progenitors of present day spirals can account
for the population of DLA absorbers, as is often assumed.
To this end, we have compared the model predictions
with several, recently determined,
statistical properties of DLAs,
in particular their number density per unit redshift,
their column density distribution, and its integral
which gives $\Omega_{\rm DLA}$, 
the total mass of neutral gas traced by 
DLAs. In addition we have brought together available
imaging data on these absorbers at $z < 1$ from the literature,
to compare with the morphologies, sizes and luminosities
of the model galaxies. Our main results can be summarised as follows.

1. The models are reasonably successful at reproducing
many of the properties of DLAs at redshifts $z \simlt 2$, 
within the uncertainties in the measurements. 
Specifically, the number density
of absorbers per unit redshift and about half
of the neutral gas mass they trace 
can be reproduced with evolving disk galaxies in our models.
Dwarf galaxies, which have not been considered here,
may account for some of the DLA absorbers, 
but this population is not the dominant one.

2. Normal spirals and low surface brightness galaxies
make comparable contributions in our models to both the
numbers of DLAs and their neutral gas mass.
Turning this statement around, it is only with
the inclusion of LSBs that it is possible to reproduce
the statistics of DLAs in our models. While LSBs may not be 
as numerous as high surface brightness galaxies, their large
dimensions and high gas content combine to make a significant
contribution to the overall cross section for H~I absorption.
The scant imaging data available at $z < 1$ is broadly     
in agreement with this conclusion, although the observed 
distribution of impact parameters $D$ seems to be narrower, 
and more peaked towards lower values of $D$, than our models predict.
The distributions of luminosities agree at the bright
end, but the data also include a few cases of very faint 
DLA-producing galaxies which are not predicted by the models. 
More uniform and extensive imaging surveys are required
to reach firm conclusions on both of these points. Furthermore, 
the properties of LSBs, which have been included in our models
according to a rather simple recipe, need to be investigated further
for a proper comparison with DLAs.

3. We have investigated the effects of a possible dust-induced 
bias in current DLA samples, which may lead to 
an underestimate of the relative number of
sightlines through galactic regions 
where the column densities
of gas and metals are both high.
We have found this potential
problem to have a relatively minor effect, particularly
on $\Omega_{\rm DLA}$, in agreement with the initial results
from the CORALS survey by Ellison et al. (2001).

4. As we move from $z \sim 2$ to higher redshifts, models 
based on extrapolating back in time the properties of
today's disk galaxies are progressively less successful
in reproducing the statistics of DLAs. 
The data can be interpreted as evidence for the 
existence of an additional population of DLAs, of generally
lower column density, which dominate the number
density of absorbers over the progenitors of today's disks
beyond $z \simeq 3$. Possibly these are sub-units destined
to merge with each other and eventually with more massive
galaxies by $z \simeq 2$. This interpretation is supported
by the failure up to now to detect the galaxies producing
DLAs at $z \simeq 4$ in deep images reaching down to 
$\simlt 1/4 L^{\ast}$.
The history of assembly of today's galactic disks may 
well be reflected in the redshift evolution of the number
density of DLAs.\\

Samuel Boissier would like to acknowledge the support 
of a Framework 5 Marie Curie fellowship through 
contract number HPMF-CT-2000-00521.

\bsp

\label{lastpage}

\end{document}